\documentclass[prc,aps,floats,twocolumn,groupedaddress,floatfix,showpacs]{revtex4}
\usepackage{graphicx,amssymb,amsmath}
\usepackage{subfigure}
\begin{document}
\title{Calculation of geometric phases in electric dipole searches with trapped spin-1/2 particles based on direct solution of the Schr\"odinger equation}

\author{A.~Steyerl}
\email{asteyerl@mail.uri.edu}
\affiliation{Department of Physics, University of Rhode Island, Kingston, RI 02881, U. S. A.}
\author{R.~Golub}
\affiliation{Physics Department, North Carolina State University, Raleigh, NC 27606, U. S. A.}
\author{C.~Kaufman}
\affiliation{Department of Physics, University of Rhode Island, Kingston, RI 02881, U. S. A.}
\author{G.~M\"uller}
\affiliation{Department of Physics, University of Rhode Island, Kingston, RI 02881, U. S. A.}
\author{S.~S.~Malik}
\affiliation{Department of Physics, University of Rhode Island, Kingston, RI 02881, U. S. A.}
\author{A.~M.~Desai}
\affiliation{Department of Physics, University of Rhode Island, Kingston, RI 02881, U. S. A.}

\begin{abstract}
Pendlebury $\textit{et al.}$ [Phys. Rev. A $\textbf{70}$, 032102 (2004)] were the first to investigate the role of geometric phases in searches for an electric dipole moment (EDM) of elementary particles based on Ramsey-separated oscillatory field magnetic resonance with trapped ultracold neutrons and comagnetometer atoms. Their work was based on the Bloch equation and later work using the density matrix corroborated the results and extended the scope to describe the dynamics of spins in general fields and in bounded geometries. We solve the Schr\"odinger equation directly for cylindrical trap geometry and obtain a full description of EDM-relevant spin behavior in general fields, including the short-time transients and vertical spin oscillation in the entire range of particle velocities. We apply this method to general macroscopic fields and to the field of a microscopic magnetic dipole.
\end{abstract}
\maketitle

%
\section{Introduction}\label{sec:I}
%
Observation of a permanent electric dipole moment (EDM) of the neutron at a level significantly higher than the Standard Model (SM) prediction of $\sim 10^{-32}$ e cm would be strong evidence for ``new physics beyond the SM''. 
While no finite EDM has been detected so far \cite{BAK01,HAR01}, the search technique which uses stored ultracold neutrons (UCN) has reached a high sensitivity of order $10^{-26}$ e cm where extreme vigilance with respect to small false effects is called for. Based on Commins' work \cite{COM01} on the geometric phase (GP) in EDM work with a thallium beam, Pendlebury {\it et al.}~\cite{PEN01} calculated the GPs accumulated by trapped UCN and cohabiting atoms serving as a magnetometer. These phases arise as a result of the motional magnetic field $\mathbf{B}$$_{v}=($$\mathbf{E}\times \mathbf{v})/c^2$ in combination with the non-zero static magnetic field inhomogeneity present in any experiment. The results of Ref.~\cite{PEN01} were obtained by integrating the Bloch equation for spin evolution in a time dependent magnetic field. A key result of this analysis is an analytical expression, Eq.~(78) of Ref.~\cite{PEN01}, for the GP mimicking an EDM for cylindrical cell and field geometry with uniform gradient and for a single particle velocity.

An identical expression was obtained in works \cite{LAM01,BAR01} based on the spin density matrix \cite{RED01,ABR01,RED02}. Aside from its fundamental relevance to NMR physics and generally to particles in bounded geometries \cite{MCG01,SWA01}, this method yields analytical results for general magnetic fields in rectangular geometries. As a test bench the authors of \cite{SWA01} considered a confined polarized gas exposed to a magnetic field with a general uniform gradient. Recently, Pignol and Roccia \cite{PIG01} showed that in the non-adiabatic limit of large particle velocity the frequency shift linear in $E$ can generally be expressed as a volume average of the field and obtained analytic results for the general gradient field as well as for the case of a microscopic magnetic dipole field. The latter had previously been shown in Ref.~\cite{HAR02} to lead to an enhancement of frequency shift relative to that for macroscopic fields and had been further analyzed for rectangular geometries in the diffusion approximation in Ref.~\cite{CLA01}.

In the present article we directly solve the Schr\"odinger equation, up to second order of perturbation, for UCNs and comagnetometer atoms like $^{199}$Hg in a Ramsey-type EDM experiment with uniform vertical electric field $\mathbf{E}$ and arbitrary small inhomogeneity of the vertical static Larmor field $\mathbf{B}_0$. We assume a cylindrical measurement cell, as for the ILL experiments \cite{BAK01,HAR01,PEN01} and in projects \cite{ALT01,FRE01}. Although our method also allows the analysis of curved paths (e.g., slightly bent due to the Coriolis force) we focus on straight-path motion between successive specular reflections on the cylinder wall and neglect spin relaxation due to gas scattering and partial diffusivity and depolarization in wall reflections. Parts of the present work have previously been presented in \cite{STE01}.

We determine the spinor evolution as a function of an arbitrary number $n$ of wall reflections during the period of free spin precession in the Ramsey scheme, whereas the previous work had been restricted to the asymptotic frequency shifts (for $n\gg 1$). Where a comparison is possible we obtain agreement with the earlier results. As further novel results of the approach we analyze the vertical spin oscillations associated with the perturbation of Larmor precession due to the field inhomogeneities.

Following the derivation of the general solution of the Schr\"odinger equation to second order in the perturbation in Sec.~\ref{sec:II} we discuss, in Sec.~\ref{sec:III}, the original model of uniform vertical field gradient and, in Sec.~\ref{sec:IV} and Appendix \ref{sec:A}, general uniform and non-uniform field gradients as more elaborate models of macroscopic fields. Section \ref{sec:V} deals with a point magnetic dipole oriented vertically as an example of a microscopic field distribution.
%
\section{General solution of the Schr\"odinger equation}\label{sec:II}
%
%
\subsection{Single chord}\label{sec:II.A}
%

Both the neutrons and the atoms used or considered as comagnetometers ($^{199}$Hg, $^3$He, $^{129}$Xe) have spin 1/2, thus their interaction with magnetic fields is described by a Hamiltonian involving the Pauli matrices $\sigma_{x}$, $\sigma_{y}$, $\sigma_{z}$ in the form (setting $\hbar=1$)
\begin{equation}\label{1}
\mathcal{H} = -\mu\boldsymbol{\sigma}\cdot{\mathbf{B}} = \frac{1}{2}\left[\begin{array}
[c]{cc}
\omega_{0} & \Sigma^{\ast}\\
\Sigma & -\omega_{0}
\end{array}\right],
\end{equation}
where the magnetic moment $\mu$ of the neutron and of $^{3}$He and $^{129}$Xe atoms is negative: $\mu/\mu_{N}=-1.913$, $-2.128$, $-0.778$ for the neutron, $^{3}$He and $^{129}$Xe (where the nuclear magneton is $\mu_{N}=0.505\times 10^{-26}$ Am$^{2}$). $^{199}$Hg has a positive moment $\mu_{\textrm{Hg}}/\mu_{N}=+0.5059$. With magnetic field $B_{0}$ in the measurement cell pointing in the $+z$ direction and the plus sign being defined by the right-hand rule around the $z$-axis, the Larmor frequency $\omega_{0}=-2\mu B_{0}$ is positive for particles with negative moment and negative for those with positive moment. 

\begin{figure}[tb]
  \begin{center}
 \includegraphics[width=77mm]{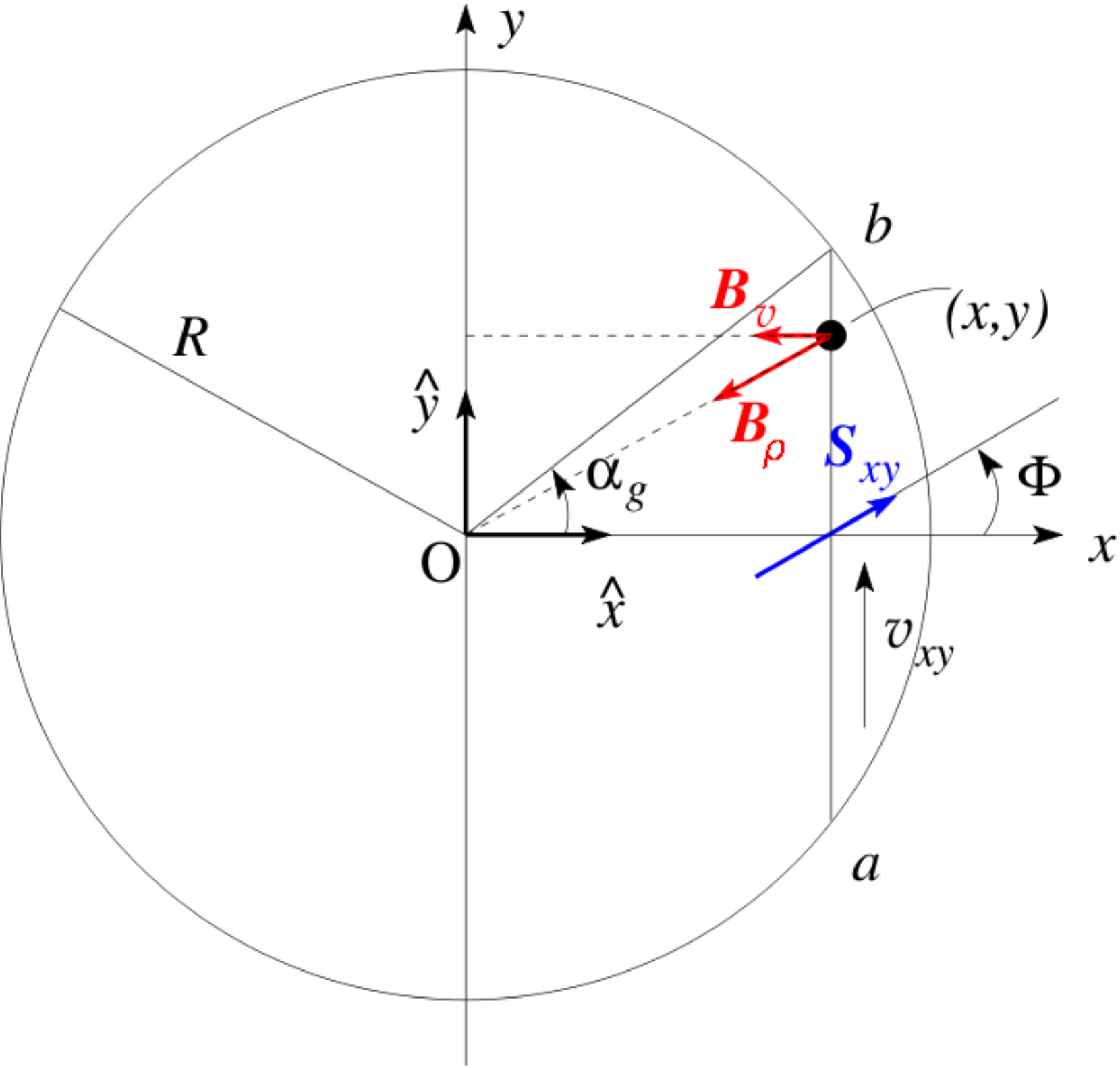}
\end{center}
\caption{(Color online) In a projection onto the horizontal ($x,y$) plane, a particle with spin projection $S_{xy}$ moves at constant velocity $v_{xy}$ in a cylindrical cell of radius $R$ along a straight path segment between successive collisions with the sidewall at $a$ and $b$. The segment is characterized by the angle $\alpha_{g}$ and for each segment we choose a coordinate system where the path is along the $y$ direction. The moving particle experiences the small horizontal magnetic fields $\mathbf{B}_{\rho}$ and $\mathbf{B}_{v}$, where $\mathbf{B}_{\rho}$ is an arbitrary small fluctuation and $\mathbf{B}_{v}$ is the motional magnetic field.}
  \label{fig:one}
\end{figure}

In (\ref{1}) the perturbing field, due to a small magnetic field inhomogeneity and to the motional field $\mathbf{E\times v}/c^{2}$ in a strong electric field $\mathbf{E}$ in the $z$-direction, is given by
\begin{equation}\label{2}
\Sigma(t)=\omega_{x}+i\omega_{y}=-2\mu(B_{x}+iB_{y}). 
\end{equation}%
For motion at constant in-plane velocity $v=v_{xy}$ along a straight path in the $y$-direction, as illustrated in Fig.~\ref{fig:one}, we can split off the $E$-dependent term $-\eta\Omega\omega_{0}$:
\begin{equation}\label{3}
\Sigma(t)=-\eta\Omega\omega_{0}+\Sigma_{B}(t),
\end{equation}  
where $\Sigma_{B}$ is the contribution of the static field inhomogeneity and $\Omega$, $\eta$ are dimensionless parameters; for velocity: $\Omega=v/(R\omega_{0})$ (with cell radius $R$); and for electric field: $\eta=B_{v}/(\Omega B_{0})=Ev/(\Omega B_{0}c^{2})=R\omega_{0}E/(B_{0} c^{2})$. 

In terms of spinor components $\alpha(t)$ and $\beta(t)$ the Schr\"odinger equation
\begin{equation}\label{4}
i\frac{d}{d t}\left[
\begin{array}
[c]{c}%
\alpha\\
\beta
\end{array}
\right]  =\frac{1}{2}\left[
\begin{array}
[c]{cc}%
\omega_{0} & \Sigma^{\ast}\\
\Sigma & -\omega_{0}%
\end{array}
\right]  \left[
\begin{array}
[c]{c}%
\alpha\\
\beta
\end{array}
\right],
\end{equation}
in the lab frame, has components
\begin{equation}\label{5}
i\dot{\alpha}=\frac{1}{2}\omega_{0}\alpha+\frac{1}{2}\Sigma^{\ast}\beta,\,\,\,i\dot{\beta}=\frac{1}{2}\Sigma\alpha-\frac{1}{2}\omega_{0}\beta.
\end{equation}%

Introducing the frame rotating with the Larmor frequency $\omega_{0}$,%
\begin{equation}\label{6}
\alpha=\alpha_{r}e^{-i\omega_{0}t/2},\,\,\,\beta =\beta_{r}e^{i\omega_{0}t/2},%
\end{equation}%
Eqs.~(\ref{5}) become
\begin{equation}\label{7}
i\dot{\alpha}_{r}=\frac{1}{2}\Sigma^{\ast}\beta_{r}e^{i\omega_{0}t},\,\,\,i\dot{\beta}_{r}=\frac{1}{2}\Sigma\alpha_{r}e^{-i\omega_{0}t}%
\end{equation}%
which we combine to give a second-order ODE for $\alpha_{r}(t)$:
\begin{equation} \label{8}
\ddot{\alpha}_{r}-\left(i\omega_{0}+ \frac{\dot{\Sigma}^{\ast}}{\Sigma^{\ast}}%
\right)  \dot{\alpha}_{r}  =-\frac{1}{4}\left|  \Sigma\right|  ^{2}%
\alpha_{r}.%
\end{equation}

The term on the right-hand side (rhs) of (\ref{8}), typically, is some eight orders of magnitude smaller than those on the lhs. We can, therefore, consider it as a small perturbation and substitute for $\alpha_{r}$ on the rhs the solutions $\alpha_{r0}$ of the unperturbed, homogeneous equation in which the rhs is replaced by zero. Solving the inhomogeneous equation thus constructed then provides the next order to $\alpha_{r}$ in an expansion in the second order quantity $\left|  \Sigma\right|  ^{2}$.

The homogeneous equation $\ddot{\alpha}_{r0}-\left(i\omega_{0}+ \dot{\Sigma}^{\ast}/\Sigma^{\ast}\right)\dot{\alpha}_{r0}=0$ is a first-order ODE for $\dot{\alpha}_{r}(t)$ with solutions $\alpha^{(1)}_{r0}=C_{1}$ and $\alpha^{(2)}_{r0}=C_{2}\Sigma^{\ast}_{i}(t)$, where $C_{1}$ and $C_{2}$ are constants and $\Sigma_{i}(t)$ is the indefinite integral
\begin{equation} \label{9}
\Sigma_{i}(t)=\int\! dt\,\textrm{e}^{-i\omega_{0}t}\Sigma(t).
\end{equation} 
If we are interested only in the frequency shift linear in $E$ we can simplify (\ref{9}) by  splitting this term off, as in (\ref{3}):
\begin{equation}\label{10}
\Sigma_{i}(t)=-i\eta\Omega\textrm{e}^{-i\omega_{0}t}+\Sigma_{iB}(t)
\end{equation}
with $\Sigma_{iB}(t)=\int\! dt\,\textrm{e}^{-i\omega_{0} t} \Sigma_{B}(t)$.

Substituting $\alpha^{(1)}_{r0}=C_{1}$ for $\alpha_{r}$ on the rhs of Eq.~(\ref{8}) we find as the solution of this first-order inhomogeneous ODE for $\dot{\alpha}_{r}(t)$:
\begin{equation} \label{11}
\dot{\alpha}^{(1)}_{r}(t)=-\frac{C_{1}}{4}\Sigma^{\ast}(t)\textrm{e}^{i\omega_{0}t}\Sigma_{i}(t).
\end{equation} 
Integrating (\ref{11}) yields, up to second order in the perturbation,
\begin{equation} \label{12}
\alpha^{(1)}_{r}(t)=C_{1}\left[1-F(t)\right]
\end{equation} 
where
\begin{equation} \label{13}
F(t)=\frac{1}{4}\int\! dt\,\Sigma^{\ast}(t)\textrm{e}^{i\omega_{0}t}\Sigma_{i}(t)
\end{equation}
represents the second-order correction. 

Next, we use the second homogeneous solution, $\alpha^{(2)}_{r0}=C_{2}\Sigma^{\ast}_{i}(t)$, on the rhs of Eq.~(\ref{8}) and could calculate the next-order correction to $\alpha^{(2)}_{r}(t)$ as the solution of this inhomogeneous equation. However, $C_{2}\Sigma^{\ast}_{i}(t)$ is already of higher order in the perturbation. Therefore, all terms up to second order are included in the solution
\begin{equation} \label{14}
\alpha_{r}(t)=\alpha^{(1)}_{r}(t)+ \alpha^{(2)}_{r}(t)=C_{1}\left[1-F(t)\right]+C_{2}\Sigma^{\ast}_{i}(t).
\end{equation}

Now we calculate $\beta_{r}$ from the first of Eqs.~(\ref{7}): 
\begin{equation}\label{15}
\beta_{r}(t) =\frac{2i\dot{\alpha}_{r}}{\Sigma^{\ast}}e^{-i\omega_{0}t}=i\left(-\frac{1}{2}C_{1}\Sigma_{i}\left(  t\right)+2C_{2}\right)
\end{equation}
and determine $C_{1}$ and $C_{2}$ from the initial conditions $\alpha_{r}(t_{0})=1$, $\beta_{r}(t_{0})=0$ for spin up at $t=t_{0}$:
\begin{equation}\label{16}
C_{1}=1-\frac{1}{4}|\Sigma_{i}(t_{0})|^{2}+F(t_{0}),\,\,\,C_{2}=\frac{1}{4}\Sigma_{i}(t_{0}),
\end{equation}
which is correct to second order.

Collecting terms from (\ref{14})-(\ref{16}) we obtain for the solution of Eq.~(\ref{8}) for initial spin up:
\begin{equation}\label{17}
\alpha_{r}(t,t_{0})=1-\Big(F(t)-F(t_{0})\Big)+\frac{1}{4}\Sigma_{i}(t_{0})\Big(\Sigma^{\ast}_{i}(t)-\Sigma^{\ast}_{i}(t_{0}) \Big),
\end{equation}
\begin{equation}\label{18}
\beta_{r}(t,t_{0})=-\frac{i}{2}\Big(\Sigma_{i}(t)-\Sigma_{i}(t_{0}) \Big),
\end{equation}
with
\begin{align}\label{19}
&\Sigma_{i}(t)-\Sigma_{i}(t_{0})=\int_{t_{0}}^{t}dt^{\prime}e^{-i\omega_{0}t^{\prime}%
}\Sigma\left(  t^{\prime}\right),\nonumber \\
&F(t)-F(t_{0})=\frac{1}{4}\int_{t_{0}}^{t}dt^{\prime}\Sigma^{\ast}\left(  t^{\prime}\right)e^{i\omega_{0}t^{\prime}}\Sigma_{i}\left(  t^{\prime}\right),\\
&\left[F(t)-F(t_{0})\right]^{E~\textrm{odd}}=\nonumber \\
&-\frac{\eta\Omega}{4}\left\{\omega_{0}\int_{t_{0}}^{t}dt^{\prime}\textrm{e}^{i\omega_{0}t^{\prime}}\Sigma_{iB}\left(  t^{\prime}\right)+i\int_{t_{0}}^{t}dt^{\prime}\Sigma^{\ast}_{B}\left(  t^{\prime}\right)\right\},\nonumber
\end{align}
from (\ref{3}), (\ref{9}), (\ref{10}) and (\ref{13}). The last expression in (\ref{19}) represents the term linear in $E$. 

The above solution is for a system that starts in the spin up state ($\alpha_{r}(t_{0})=1$). Combining with the solution where the system starts in the spin down state ($\beta_{r}(t_{0})=1$) we obtain the general spinor solution in terms of a matrix $M_{r}(t,t_{0})$:
\begin{equation}\label{20}
\psi_{r}\left(  t\right)  =\left[
\begin{array}
[c]{c}%
a_{r}\left(  t\right) \\
b_{r}\left(  t\right)
\end{array}
\right]  = M_{r}(t,t_{0})\left[
\begin{array}
[c]{c}%
a_{r}\left(  t_{0}\right) \\
b_{r}\left(  t_{0}\right)
\end{array}
\right],
\end{equation}
\begin{equation}\label{21}
M_{r}(t,t_{0})=\left[
\begin{array}
[c]{cc}%
\alpha_{r}\left(  t,t_{0}\right)  & -\beta_{r}^{\ast}\left(  t,t_{0}\right) \\
\beta_{r}\left(  t,t_{0}\right)  & \alpha_{r}^{\ast}\left(  t,t_{0}\right)
\end{array}
\right].
\end{equation}
It describes the evolution along the chord in the rotating system. Transforming back to the lab system we obtain%
\begin{align}
&M(t,t_{0})=\left[\!
\begin{array}
[c]{cc}%
e^{-i\omega_{0}t/2} & 0\\
0 & e^{i\omega_{0}t/2}%
\end{array}\!
\right] M_{r}(t,t_{0}),\label{22}\\
&\psi\left(  t\right)  =M(t,t_{0}) \left[
\begin{array}
[c]{c}%
a_{r}\left(  t_{0}\right) \\
b_{r}\left(  t_{0}\right)
\end{array}
\right].\label{23}
\end{align}
The matrices $M(t,t_{0})$, $M_{r}(t,t_{0})$ are unitary given that $\alpha_{r},\beta_{r}$ are normalized.

We will now consider the start of free precession in the Ramsey scheme, i.e. the time immediately following the first $\pi/2$ pulse, and measure the time elapsed in free precession by $t_{p}$. Let the initial spin state at $t_{p}=0$ be of the general form
\begin{equation}\label{24}
\left[
\begin{array}
[c]{c}%
a_{r}\left(  t_{0}\right) \\
b_{r}\left(  t_{0}\right)
\end{array}
\right]= \left[
\begin{array}
[c]{c}%
c \\
s\textrm{e}^{i\Phi}
\end{array}
\right].
\end{equation}
For the experimentally common situation the polar spin angle equals $\theta=\pi/2$ at $t_{p}=0$, thus $c=\cos(\theta/2)=s=\sin(\theta/2)=1/\sqrt{2}$. But as the particle moves along consecutive path segments $\theta$ oscillates about $\pi/2$ due to the perturbation (\cite{PEN01} and see Sec.~\ref{sec:II.D} below).

As for the azimuthal angle $\Phi$, all spins point in the same direction in the lab frame at $t_{p}=0$. This implies that, measured relative to the random direction of particle motion at this time, $\Phi$ is uniformly distributed over all angles from $-\pi$ to $+\pi$.

To analyze motion along the initial chord we write, using (\ref{20}), (\ref{24}),
\begin{align}\label{25}
&\psi_{r}\left(  t\right)=\left[
\begin{array}
[c]{c}%
a_{r}\left(  t\right) \\
b_{r}\left(  t\right)
\end{array}
\right]=\left[
\begin{array}
[c]{cc}%
\alpha_{r}\left(  t,t_{0}\right)  & -\beta_{r}^{\ast}\left(  t,t_{0}\right) \\
\beta_{r}\left(  t,t_{0}\right)  & \alpha_{r}^{\ast}\left(  t,t_{0}\right)
\end{array}
\right]\left[
\begin{array}
[c]{c}%
c \\
s\textrm{e}^{i\Phi}
\end{array}
\right]\nonumber \\
&=\left[
\begin{array}
[c]{c}%
c\alpha_{r}(t,t_{0})-s\textrm{e}^{i\Phi}\beta^{\ast}_{r}(t,t_{0}) \\
c\beta_{r}(t,t_{0})+s\textrm{e}^{i\Phi}\alpha^{\ast}_{r}(t,t_{0})
\end{array}
\right].
\end{align}
The frequency shifts are determined by the azimuthal angle $\varphi(t)$ of spinor $\psi_{r}(t)$:
\begin{align}\label{26}
&\varphi(t)=-\arg{\left[\frac{a_{r}(t)}{b_{r}(t)}\right]}=-\arg{\left[\frac{c\alpha_{r}-s\textrm{e}^{i\Phi}\beta^{\ast}_{r}}{\textrm{e}^{i\Phi}\left(c\textrm{e}^{-i\Phi}\beta_{r}+s\alpha^{\ast}_{r} \right)}\right]}\nonumber \\
& =-\arg{\left[\frac{\left(c\alpha_{r}-s\textrm{e}^{i\Phi}\beta^{\ast}_{r}\right)\left(c\textrm{e}^{i\Phi}\beta^{\ast}_{r}+s\alpha_{r}\right) }{\textrm{e}^{i\Phi}\left|c\textrm{e}^{-i\Phi}\beta_{r}+s\alpha^{\ast}_{r} \right|^{2}}\right]}.
\end{align}
Thus the phase shift, measured in the rotating system, becomes
\begin{align}\label{27}
&\delta\varphi=\varphi-\Phi\nonumber \\
&=-\arg\left(c s \alpha^{2}_{r}+(c^{2}-s^{2})\alpha_{r}\beta^{\ast}_{r}\textrm{e}^{i\Phi}-c s \beta^{\ast 2}_{r}\textrm{e}^{2 i\Phi} \right).
\end{align}
Averaging over $\Phi$ yields
\begin{equation}\label{28}
\langle\delta\varphi(t,t_{0})\rangle=-\arg\alpha^{2}_{r}(t,t_{0})=2\arg\alpha^{\ast}_{r}(t,t_{0}),
\end{equation}
given that $\langle\Phi\rangle=0$ and all terms $\propto\!\textrm{e}^{i\Phi}$, $\textrm{e}^{2i\Phi}$, etc., average to zero.

Now we introduce dimensionless time $\tau=\omega_{0}t$, setting $\tau=0$ at the center of any chord, and evaluate the phase shift (\ref{28}) over the full initial chord from point $a$ to point $b$ of Fig.~\ref{fig:one}, i.e. for start at time $\tau=-\delta$ and end at $\tau=+\delta$ where $\delta=(\sin\alpha_{g})/\Omega$. We call this shift $\langle\delta\varphi_{0\rightarrow 1}\rangle$ and obtain from (\ref{17}):
\begin{align}\label{29}
&\langle\delta\varphi_{0\rightarrow 1}\rangle=2\nu(\delta,-\delta)\nonumber \\
&=2\operatorname{Im}\Big[F(\delta)-F(-\delta)-\frac{1}{4}\Sigma^{\ast}_{i}(\delta)\Sigma_{i}(-\delta)\Big]
\end{align}
where we have defined $\nu(\delta,-\delta)=-\arg[\alpha_{r}(\delta,-\delta)]$.

Note that expression (\ref{29}) is valid for any initial polar spin angle $\theta$, not only for $\theta=\pi/2$, and this independence will be seen to hold also for the phase shift $\langle\delta\varphi_{0\rightarrow n}\rangle$ over an arbitrary number $n$ of consecutive chords. Similarly, the asymptotic frequency shift, for $n\rightarrow \infty$, will be independent of starting point on the initial chord (which, in the experimental situation, is uniformly distributed over the chord length); thus our choice $\tau_{0}=-\delta$ made above does not narrow the scope of the analysis.
%
\subsection{Consecutive chords}\label{sec:II.B}
%
We now turn to consecutive path segments $n=2$, $3$, ...  At each wall reflection, assumed specular, the flight path is redirected by the angle $2\alpha_{g}$ and the spinor remains unchanged since the reflection time is much shorter than the Larmor period. Thus, at the end of chord $2$ we have
\begin{equation}\label{30}
\psi^{(2)}_{r}(\delta)=\left[
\begin{array}
[c]{c}%
a^{(2)}(\delta)\\
b^{(2)}(\delta)
\end{array}
\right]=M^{(2)}_{r}(\delta,-\delta)\left[
\begin{array}
[c]{c}%
c\\
s\textrm{e}^{i\Phi}
\end{array}
\right]
\end{equation}
with
\begin{equation}\label{31}
M^{(2)}_{r}(\delta,-\delta)=M_{T}(\delta,-\delta)R(\alpha_{g})M_{T}(\delta,-\delta),
\end{equation}
where
\begin{equation}\label{32}
M_{T}(\delta,-\delta)=T(-\delta)M_{r}(\delta,-\delta)T(-\delta)
\end{equation}
includes the transition from the lab to the rotating system and back via
\begin{equation}
T(-\delta)=\left[
\begin{array}
[c]{cc}%
\textrm{e}^{-i\delta/2}  & 0 \\
0  & \textrm{e}^{i\delta/2}
\end{array}
\right],\nonumber
\end{equation}
and
\begin{equation}
R(\alpha_{g})=\left[
\begin{array}
[c]{cc}%
\textrm{e}^{i\alpha_{g}}  & 0 \\
0  & \textrm{e}^{-i\alpha_{g}}
\end{array}
\right]\nonumber
\end{equation}
is the transformation matrix for angular change $2\alpha_{g}$ to the coordinate system of the next chord (with $y$-axis in direction of motion).

Extending the analysis to $n\ge 3$ consecutive segments we generalize (\ref{30}) to the recursion relation
\begin{align}\label{33}
&\psi^{(n)}_{r}(\delta)=\left[
\begin{array}
[c]{c}%
a^{(n)}(\delta)\\
b^{(n)}(\delta)
\end{array}
\right]=M^{(n)}_{r}(\delta,-\delta)\left[\begin{array}{c}
c  \\
s\textrm{e}^{i\Phi}
 \end{array}\right],\nonumber  \\
&M^{(n)}_{r}(\delta,-\delta)=M_{T}(\delta,-\delta)R(\alpha_{g})M^{(n-1)}_{r}(\delta,-\delta), 
\end{align}
where $M^{(2)}_{r}(\delta,-\delta)$ is given in (\ref{31}).

Performing the matrix multiplications (\ref{33}) repeatedly we obtain a sequence of matrices whose general form can be deduced from the first few terms, say for $n=2$ up to $5$. For any $n$, the result is a unitary transfer matrix of form
\begin{equation}\label{34}
M^{(n)}_{r}(\delta,-\delta)=\left[
\begin{array}
[c]{cc}%
g_{r}  & -h^{\ast}_{r} \\
h_{r}  & g^{\ast}_{r}
\end{array}
\right],
\end{equation}
and applying the same algebra to $g_{r}$, $h_{r}$ as to $\alpha_{r}$, $\beta_{r}$ in (\ref{26}) we can show that the overall phase advance up to the end of chord $n$, relative to the rotating frame and averaged over initial $\Phi$, is
\begin{align}\label{35}
&\langle\delta\varphi_{0\rightarrow n}\rangle=-2\arg g_{r}\nonumber \\
&=2n\nu(\delta,-\delta)-2 s^{2}_{\mu}(\delta,-\delta)\sum_{k=1}^{n-1}(n-k)\sin[2k(\alpha_{g}-\delta)]\nonumber \\
&=2n\nu(\delta,-\delta)-n s^{2}_{\mu}(\delta,-\delta)\cot(\alpha_{g}-\delta)\\
&+\frac{1}{2}s^{2}_{\mu}(\delta,-\delta)\frac{\sin[2n(\alpha_{g}-\delta)]}{\sin^{2}(\alpha_{g}-\delta)}, \nonumber
\end{align}
where $s_{\mu}(\delta,-\delta)=|\beta_{r}(\delta,-\delta)|$ [with $\beta_{r}$ from (\ref{18})] and we have used summation relations (\cite{GR01}, 1.341.1, 1.352) for the sines of multiples of an angle. The first term in (\ref{35}), $2n\nu(\delta,-\delta)$, is the sum of single-chord contributions (\ref{29}) and the remainder, $\propto\! s^{2}_{\mu}(\delta,-\delta)$, ensures that the spinor remains unchanged at wall reflections.
 
%
\subsection{Frequency shifts}\label{sec:II.C}
%
Dividing the phase shift (\ref{35}) by the net elapsed time $2 n\delta/\omega_{0}$ we obtain the frequency shift, for arbitrary $n\ge 2$:
\begin{align}\label{36}
&\frac{(\delta\omega)_{n}}{\omega_{0}}=\frac{\langle\delta\varphi_{0\rightarrow n}\rangle}{2 n\delta}=\nonumber \\
&\frac{\nu(\delta,-\delta)}{\delta}-\frac{s^{2}_{\mu}(\delta,-\delta)}{2\delta}\cot(\alpha_{g}-\delta)\\
&+\frac{s^{2}_{\mu}(\delta,-\delta)}{4 n\delta}\frac{\sin[2 n(\alpha_{g}-\delta)]}{\sin^{2}(\alpha_{g}-\delta)}.\nonumber
\end{align}
For large $n$ the last term in (\ref{36}) becomes negligible and the asymptotic shift is
\begin{equation}\label{37}
\frac{(\delta\omega)_{n \gg 1}}{\omega_{0}}=\frac{\nu(\delta,-\delta)}{\delta}-\frac{s^{2}_{\mu}(\delta,-\delta)}{2\delta}\cot(\alpha_{g}-\delta).
\end{equation}
%
\subsection{Vertical spin oscillation}\label{sec:II.D}
%
The change of polar angle $\theta$, in form of vertical spin oscillations, acts as the source of azimuthal phase shifts. We will deduce the properties of these oscillations from the explicit form of transfer matrix (\ref{34}), as determined by the matrix multiplications in (\ref{33}). Assuming start at $t_{p}=0$ at $\theta=\pi/2$, $\Phi=0$ we write
\begin{equation}\label{38}
\psi^{(n)}_{r}(\delta)=\left[
\begin{array}
[c]{c}%
a^{(n)}(\delta) \\
b^{(n)}(\delta)
\end{array}
\right]=\frac{1}{\sqrt{2}}\left[
\begin{array}
[c]{cc}%
g_{r}  & -h^{\ast}_{r} \\
h_{r}  & g^{\ast}_{r}
\end{array}
\right]\left[
\begin{array}
[c]{c}%
1 \\
1
\end{array}
\right]
\end{equation} 
and derive the following analytic expression for the polar spin angle $\theta^{(n)}(\delta)$ at the end of chord $n$ from the explicit form of $g_{r}$, $h_{r}$:
\begin{align}\label{39}
&\cos\theta^{(n)}(\delta)\nonumber \\
&=\cos^{2}(\theta^{(n)}/2)-\sin^{2}(\theta^{(n)}/2)=|a^{(n)}|^{2}-|b^{(n)}|^{2}\nonumber\\
&=2 s_{\mu}(\delta,-\delta)\frac{\sin[n(\alpha_{g}-\delta)]}{\sin(\alpha_{g}-\delta)}\times\nonumber \\
&\cos\Big\{(n-1)(\alpha_{g}-\delta)-\delta+\arg[\beta_{r}(\delta,-\delta)] \Big\},
\end{align}
for arbitrary $n\ge 2$. We have again used sum rules (\cite{GR01}, 1.341.1, 1.352).

\section{Example of uniform vertical $B_{z}$ gradient and cylindrical geometry}\label{sec:III}

Referring to the geometry shown in Fig.~\ref{fig:one} we parametrize a uniform vertical static field gradient $\partial B_{z}/\partial z$ by $\zeta=(R/2 B_{0})(\partial B_{z}/\partial z)$, thus the horizontal static magnetic field is $\mathbf{B}_{\rho}=-(\boldsymbol{\rho}/2)(\partial B_{z}/\partial z)=-\zeta B_{0}\boldsymbol{\rho}/R$ with $\boldsymbol{\rho}=(x,y)$. An estimate typical of the ILL experiments \cite{BAK01,HAR01,PEN01} is $\zeta\simeq\!1.2\times 10^{-4}$ for $B_{0}=1\mu$T and $|\langle\partial B_{z}/\partial z\rangle|=1$ nT/m. These values are averages over a measuring cell with radius $R=0.235$ m and height $H=0.12$ m.

\begin{figure}[tb]
  \begin{center} 
 \includegraphics[width=43mm]{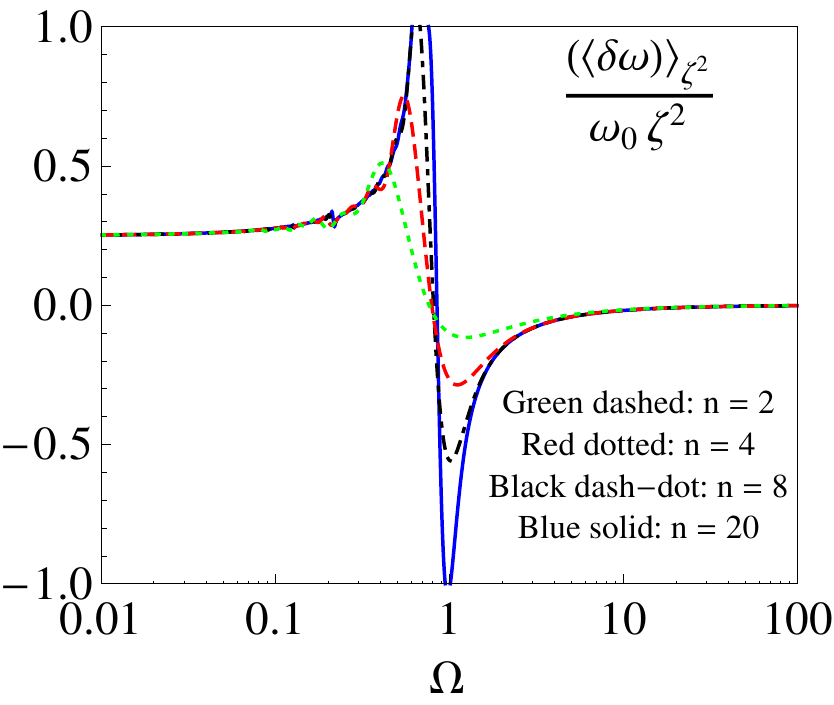}\hspace{0mm}%
 \includegraphics[width=43mm]{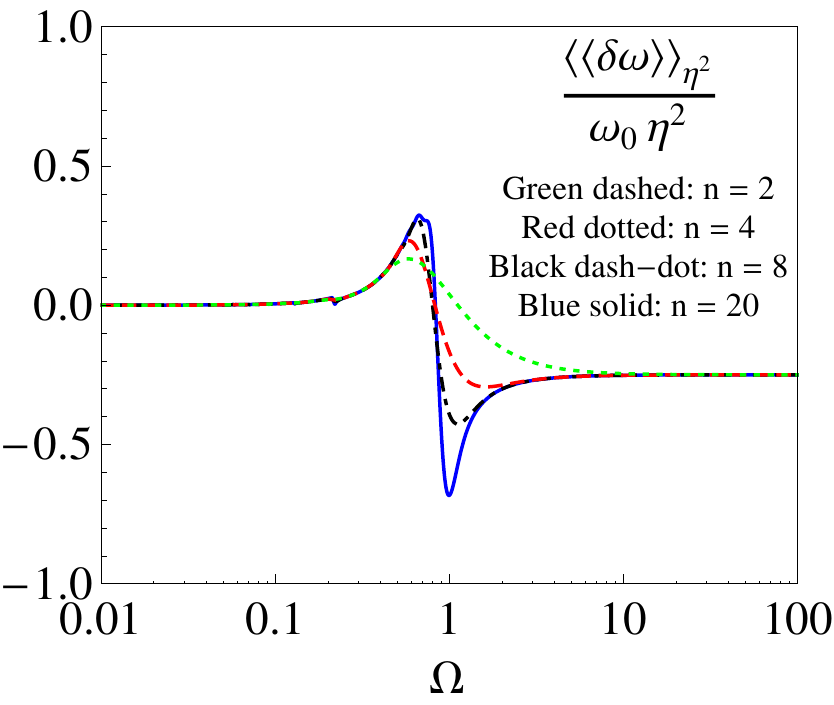}
  \includegraphics[width=43mm]{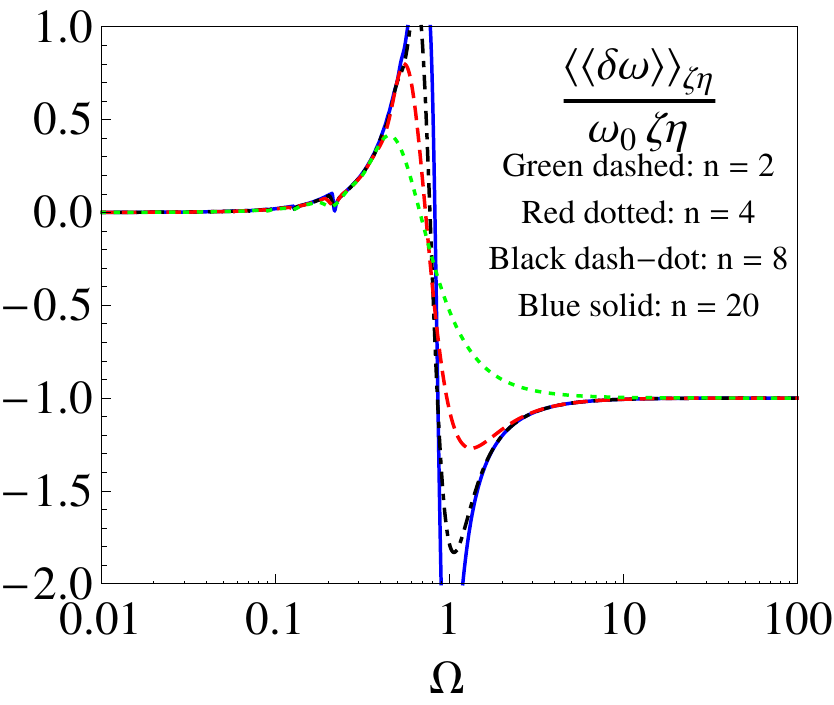}\hspace{0mm}%
  \includegraphics[width=43mm]{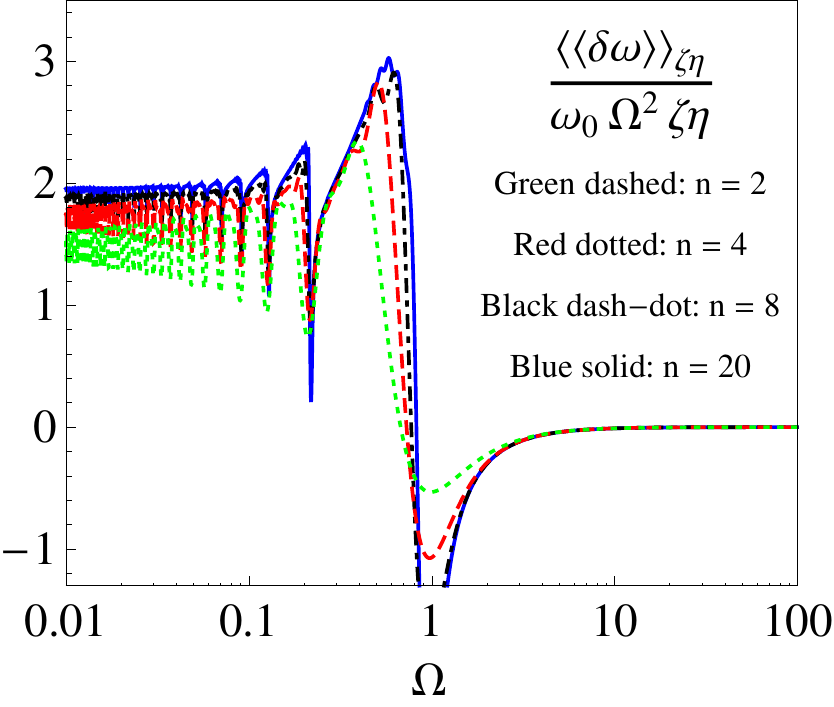}
\end{center}
\caption{(Color online) For the uniform vertical gradient field we plot the normalized mean frequency shifts from Eq.~(\ref{36}) against $\Omega=v/(R\omega_{0})$. The transients for $2$, $4$, and $8$ chords are compared with the asymptotic shifts which are represented here by $n=20$ within $\simeq\!2\%$. The latter are obtained from Eqs.~(\ref{43}-\ref{45}) by averaging over $\alpha_{g}$. Top left: Second-order magnetic gradient term $\langle\langle\delta\omega\rangle\rangle_{\zeta^{2}}/(\zeta^{2}\omega_{0})$. Top right: Second-order motional field term $\langle\langle\delta\omega\rangle\rangle_{\eta^{2}}/(\eta^{2}\omega_{0})$. Bottom left: frequency shift linear in $E$, $\langle\langle\delta\omega\rangle\rangle_{\zeta\eta}/(\zeta\eta\omega_{0})$; Bottom right: the same but divided by $\Omega^{2}$ to show the adiabatic limit $\Omega \rightarrow 0$. For $n\rightarrow \infty$ this limit is $2.0$.}
  \label{fig:two}
\end{figure}

Adding the motional magnetic field $\mathbf{B}_{v}=(\mathbf{E\times v})/c^{2}$ we have, from (\ref{2}),
\begin{align}\label{40}
&\Sigma(\tau)=\omega_{x}+i\omega_{y}=\omega_{0}\frac{B_{x}+iB_{y}}{B_{0}}\nonumber \\
&=-\omega_{0}\left(\zeta\frac{x+iy}{R}+\eta\Omega\right)=-\omega_{0}\zeta\Omega(u+i\tau)
\end{align}
and from (\ref{9}),
\begin{equation}\label{41}
\Sigma_{i}(\tau)=-i\zeta\Omega\textrm{e}^{-i\tau}\left(u+1+i\tau \right)
\end{equation} 
where $u=(\eta/\zeta)+(\cos\alpha_{g})/\Omega$ and, since time is reset at the segment center: $y=vt=R\Omega\tau$.

Proceeding as in (\ref{17})-(\ref{19}), (\ref{29}), (\ref{36}) we obtain
\begin{align}\label{42}
&\beta_{r}(\delta,-\delta)=i\zeta\Omega\left(u_{1}\sin\delta -\delta\cos\delta\right),\nonumber \\
&\frac{\nu(\delta,-\delta)}{\delta}=\frac{\zeta^{2}\Omega^{2}}{6}\Big\{3u^{2}_{1}+\delta^{2}\\
&-3\frac{\sin\delta}{\delta}\Big[(u^{2}_{1}-\delta^{2})\cos\delta +2u_{1}\delta\sin\delta\Big] \Big\},\nonumber
\end{align} 
with $u_{1}=u+1$.

Using these expressions in (\ref{37}) we reproduce the asymptotic frequency shifts ($n\rightarrow \infty$) first derived in \cite{PEN01}: the second-order gradient shift ($\propto\!\zeta^{2}$, Eq.~(71) of \cite{PEN01})
\begin{align}\label{43}
&\frac{\langle\delta\omega\rangle_{\zeta^{2}}}{\omega_{0}}=\frac{\zeta^{2}}{6}\Big[(3-2\sin^{2}\alpha_{g} +3\sin^{2} \alpha_{g} \frac{\delta-\tan \delta}{\delta^{2}\tan\delta}+\nonumber \\
  &3\Big(\!\cos\alpha_{g}+\sin\alpha_{g} \frac {\tan\delta-\delta}{\delta \tan\delta}\Big)^{2}  \frac{\sin\alpha_{g}}{\sin(\delta-\alpha_{g})} \frac{\sin\delta}{\delta} \Big];
\end{align}
the second-order geometric motional shift averaged over forward/backward motion ($\propto\!\eta^{2}$, Eq.~(80) of \cite{PEN01}),
\begin{equation}\label{44}
\frac{\langle\delta\omega\rangle_{\eta^{2}}}{\omega_{0}}=\frac{\eta^{2}\Omega^{2}}{2}\Big[1+\frac{\sin^2\alpha_{g}\sin 2\delta}{2\delta\sin(\delta-\alpha_{g})\sin(\delta+\alpha_{g})}\Big];
\end{equation}
and the EDM-mimicking geometric phase shift for $E$-field reversal ($\propto\!\zeta\eta$, corresponding to Eqs.~(77), (78) in \cite{PEN01} and to Eq.~(26) in \cite{BAR01}):
\begin{equation}\label{45}
\frac{\langle\delta\omega\rangle_{\zeta \eta}}{\omega_{0}}=2\zeta \eta\Omega^{2}\Big[1+\frac{\sin^2\alpha_{g}\sin 2\delta}{2\delta\sin(\delta-\alpha_{g})\sin(\delta+\alpha_{g})}\Big].
\end{equation} 

Figure \ref{fig:two} shows the averages over an ensemble of orbits, weighted with the probability $P(\alpha_{g})=(4/\pi)\sin^{2}\alpha_{g}$ \cite{PEN01} for segment angle $\alpha_{g}$. The plots also include the transients for finite $n$ from (\ref{36}). They show that, off resonance, the asymptotic shifts are reached, within $\simeq\!2\%$, after some $n=20$ to $50$ reflections whereas this process is much slower near the resonances, as expected. For the ILL data, both the UCN range $\Omega\simeq\!0.05$ and the $^{199}$Hg range $\Omega\simeq\!20$ are quite far from the dominant resonance at $\Omega\simeq\!1$.

\begin{figure}[tb]
  \begin{center} 
 \includegraphics[width=43mm]{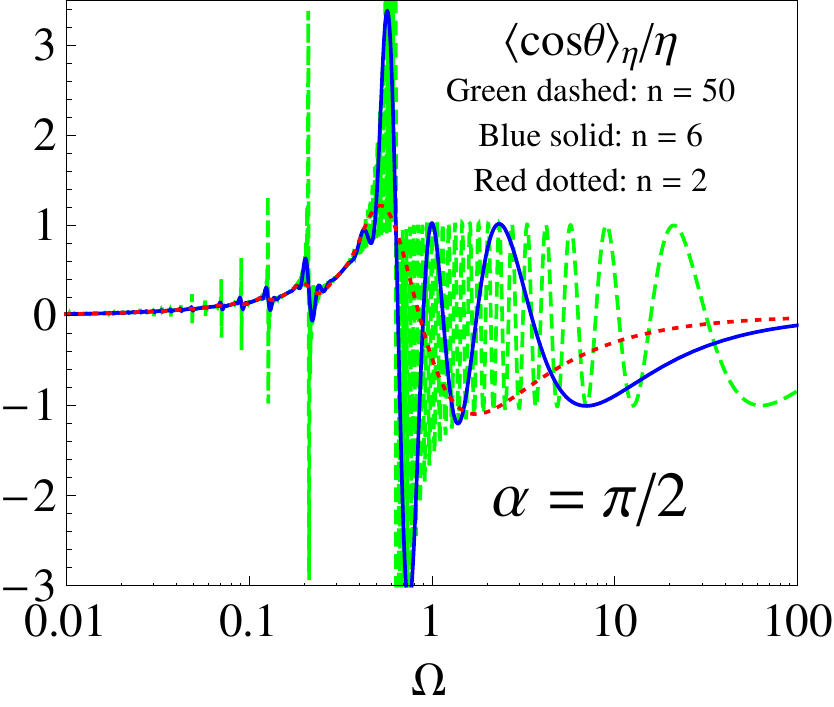}\hspace{0mm}%
 \includegraphics[width=43mm]{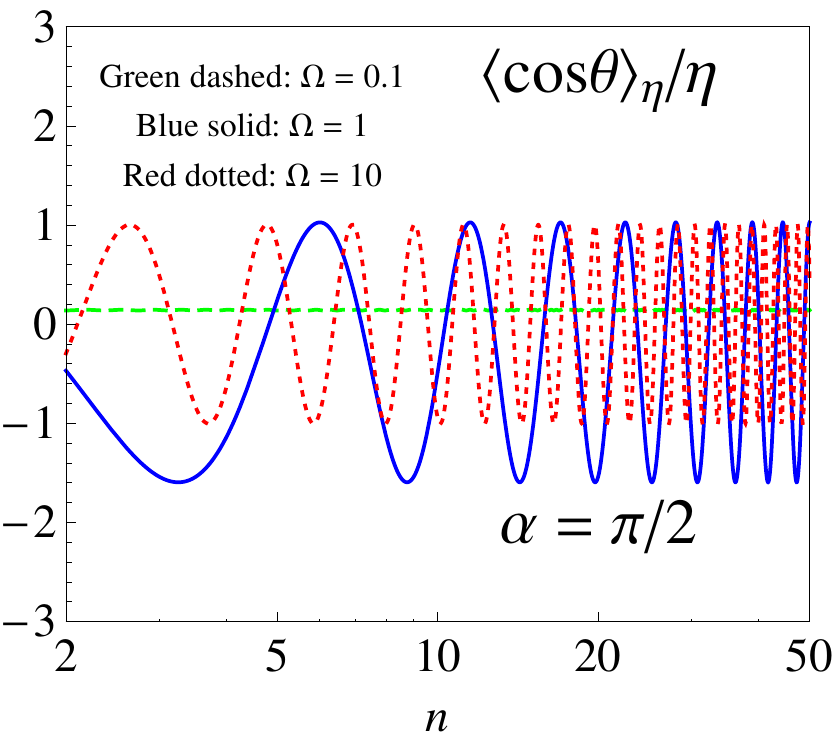}
  \includegraphics[width=43mm]{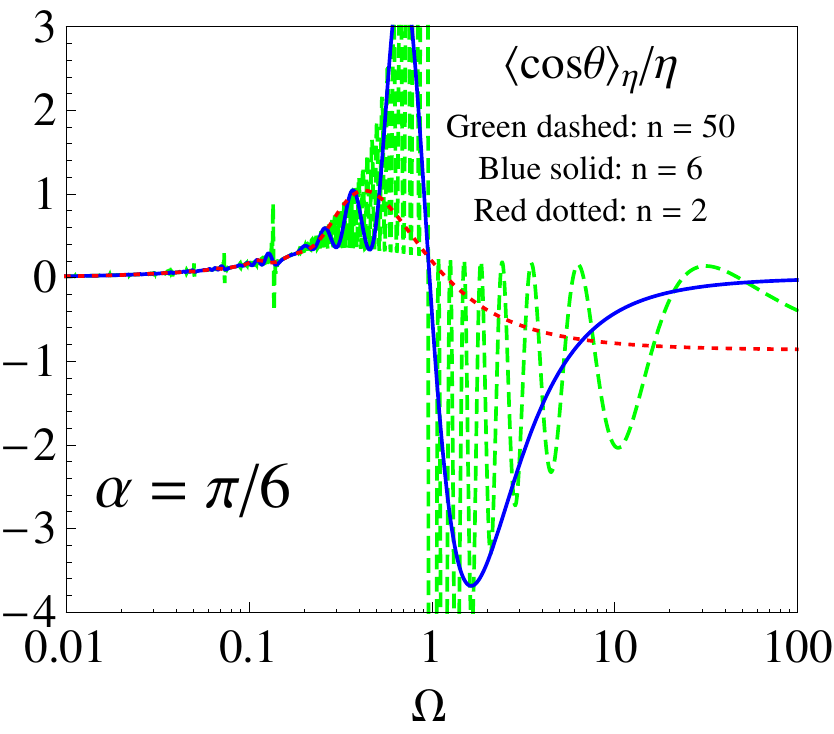}\hspace{0mm}%
  \includegraphics[width=43mm]{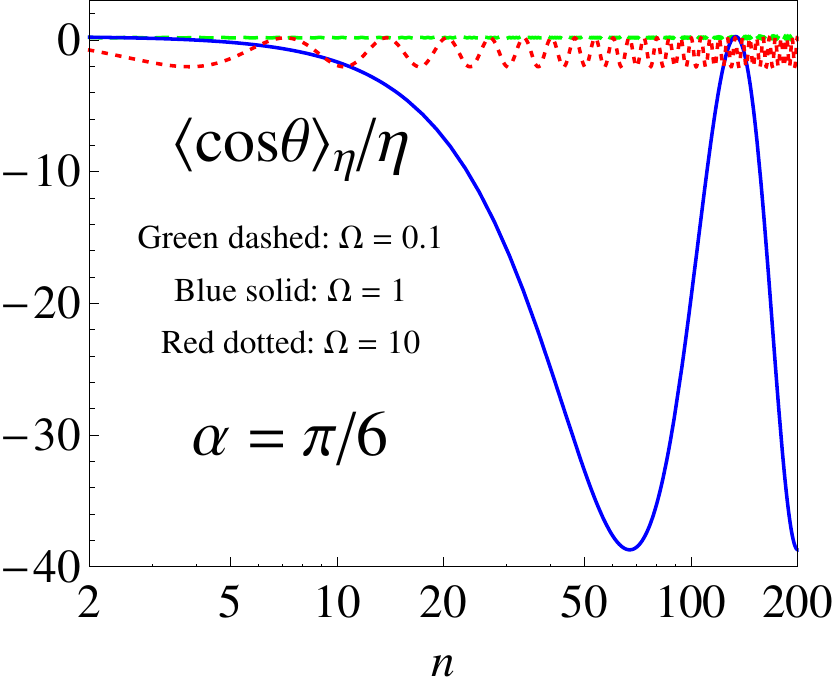}
\end{center}
\caption{(Color online) Variation of polar spin angle $\theta$ for the uniform vertical gradient field, for segment angles $\alpha_{g}=90^{\circ}$ and $30^{\circ}$. On the left, $\langle\cos\theta\rangle_{\eta}/\eta$ is plotted versus $\Omega$ for chord sequences of length $n=2$, $6$ and $50$; on the right as a function of $n$ for $\Omega=0.1$, $1$ and $10$. Proximity to a resonance is characterized by large amplitudes and long periods of oscillation.}
  \label{fig:three}
\end{figure}

The vertical spin oscillations given by (\ref{39}) have contributions $\propto\!\zeta$ and $\propto\!\eta$. Extending (\ref{39}) we have averaged $\cos\theta^{(n)}$ over all starting points on chord $n=1$, from $\tau_{0}=-\delta$ to $+\delta$, as in the experimental situation, and plot the term $\propto\!\eta$ in Fig.~\ref{fig:three} for $\alpha_{g}=\pi/2$ and $\alpha_{g}=\pi/6$ in two ways. The panels on the left show $\langle\cos\theta\rangle_{\eta}/\eta$ vs.~$\Omega$ for $n=2$, $6$ and $50$, and those on the right vs.~$n$ for $\Omega=0.1$, $1$ and $10$. As expected, the oscillation amplitude and period increase strongly near the principal resonance at $\Omega\sim\!1$.

\section{Generalized uniform field gradient}\label{sec:IV}

As a model for large-scale magnetic inhomogeneities the authors of \cite{PIG01} considered a "general uniform gradient" field
\begin{align}\label{46}
&B_{x}=G_{x}x+Q_{y}z+Q_{z}y,\nonumber \\
&B_{y}=G_{y}y+Q_{x}z+Q_{z}x
\end{align} 
derived via $\mathbf{B}=\boldsymbol{\nabla}\chi$ from the second-order polynomial for the magnetic potential,
\begin{align}\label{47}
&\chi(x,y,z)=B_{0}z+\frac{G_{x}}{2}x^2+\frac{G_{y}}{2}y^{2}\nonumber \\
&-\frac{1}{2}(G_{x}+G_{y})z^{2}+Q_{x}yz+Q_{y}zx+Q_{z}xy,
\end{align} 
which is subject to the Laplace equation $\nabla^{2}\chi=0$.

The G's and Q's are constant parameters which can be used to fit a magnetic map in an EDM measurement cell in a way more general than the uniform cylindrical vertical gradient pioneered in \cite{PEN01}. It turned out that with the more complex model the non-adiabatic limit ($\Omega\rightarrow\infty$) of the shift linear in $E$ is determined solely by the volume-averaged vertical gradient $\langle\partial B_{z}/\partial z\rangle=-(G_{x}+G_{y})$, just as for the original model. 

We will solve the Schr\"odinger equation to analyze the shift linear in $E$ for the general gradient field (\ref{46}) over the entire range of $\Omega$. To take into account the lack of cylindrical symmetry of the field $\mathbf{B}_{\rho}$ we have to average over the angle $\xi$ of a chord with given $\alpha_{g}$, relative to the static field which is given in terms of the coordinates $x,y,z$ of Fig.~\ref{fig:one}. At position $x$, $y$ of a point on a chord that is rotated about the $z$-axis by the angle $-\xi$ relative to that shown in Fig.~\ref{fig:one}, the coordinates along the rotated chord are
\begin{align}\label{48}
&x^{\prime}(\alpha_{g},\tau)=x(\alpha_{g})\cos\xi+y(\tau)\sin\xi,\nonumber \\
&y^{\prime}(\alpha_{g},\tau)=-x(\alpha_{g})\sin\xi+y(\tau)\cos\xi,\\
&z^{\prime}=z, \nonumber
\end{align}
with $x(\alpha_{g})=R\cos\alpha_{g}$, $y(\tau)=R\Omega\tau$. At this position the field is
\begin{align}\label{49}
B_{x}(\alpha_{g},\tau)&=G_{x}x^{\prime}(\alpha_{g},\tau)+Q_{y}z+Q_{z}y^{\prime}(\alpha_{g},\tau),\nonumber \\
B_{y}(\alpha_{g},\tau)&=G_{y}y^{\prime}(\alpha_{g},\tau)+Q_{x}z+Q_{z}x^{\prime}(\alpha_{g},\tau),\\
B_{z}(\alpha_{g},\tau)&=B_{0}-z(G_{x}+G_{y})+Q_{x}y^{\prime}(\alpha_{g},\tau)\nonumber \\
&+Q_{y}x^{\prime}(\alpha_{g},\tau),\nonumber\\
\langle\partial B_{z}/\partial z\rangle&=-(G_{x}+G_{y}),\nonumber
\end{align}
and rotated back by the angle $+\xi$ we have the field as seen by the particle:
\begin{align}\label{50}
&B^{\prime}_{x}(\alpha_{g},\tau)=B_{x}(\alpha_{g},\tau)\cos\xi -B_{y}(\alpha_{g},\tau)\sin\xi,\nonumber \\
&B^{\prime}_{y}(\alpha_{g},\tau)=B_{x}(\alpha_{g},\tau)\sin\xi +B_{y}(\alpha_{g},\tau)\cos\xi\nonumber \\
&B^{\prime}_{z}(\alpha_{g},\tau)=B_{z}(\alpha_{g},\tau).
\end{align}
The function $\Sigma_{B}(\alpha_{g},\tau)$, the part of $\Sigma(\tau)$ left when $-\omega_{0}\eta\Omega$ is removed as in (\ref{3}), is
\begin{equation}\label{51}
\Sigma_{B}(\alpha_{g},\tau)=\frac{\omega_{0}}{B_{0}}\left(B^{\prime}_{x}(\alpha_{g},\tau)+i B^{\prime}_{y}(\alpha_{g},\tau)\right),
\end{equation}
and the mean values of $\Sigma_{B}$ and $\Sigma_{iB}$, averaged over elevation $z$ and over a uniform distribution of angles $\xi$ from $-\pi$ to $+\pi$, are readily calculated using $\langle \cos^{2}\xi\rangle=\langle \sin^{2}\xi\rangle=1/2$, $\langle \cos\xi \rangle=\langle \sin\xi\rangle=\langle \cos\xi \sin\xi\rangle=0$ and $\langle z \rangle =0$ (with $z$ measured from the central plane), with the result
\begin{equation}\label{52}
\langle \Sigma_{B}(\alpha_{g},\tau)\rangle_{\xi,z}=\omega_{0}\frac{R}{2B_{0}}(G_{x}+G_{y})(\cos\alpha_{g}+i\Omega\tau),
\end{equation}
\begin{align}\label{53}
&\langle\Sigma_{iB}(\alpha_{g},\tau)\rangle_{\xi,z}=\nonumber \\
&\frac{iR}{2B_{0}}\textrm{e}^{-i\tau}(G_{x}+G_{y})\left(\cos\alpha_{g}+\Omega(1+i\tau)\right).
\end{align}

Proceeding as in sections ~\ref{sec:II.B}, \ref{sec:II.C} we note that all expressions required for the $E$-odd frequency shift, including those parts of $\nu(\delta,-\delta)$ and $s_{\mu}^{2}(\delta,-\delta)$ which are proportional to $\eta$, depend on $\Sigma_{B}$ and/or $\Sigma_{iB}$ linearly since products like $\Sigma^{\ast}\Sigma_{i}$ can be expanded as in the last line of Eq.~(\ref{19}). Therefore, we can take into account the random orientation of the initial chord as well as the angular changes for consecutive chords, whose directions cover the full range of angles $\xi$, by the use of averages (\ref{52}), (\ref{53}) for $\Sigma_{B}$ and $\Sigma_{iB}$. This results in the asymptotic geometric frequency shift for $E$-field reversal
\begin{align}\label{54}
&\left\langle\frac{\delta\omega}{\omega_{0}} \right\rangle^{E\rightarrow -E}=\\
&-\frac{\eta R\Omega^{2}}{B_{0}}(G_{x}+G_{y})\Big[1+\frac{\sin^{2}\alpha_{g}\sin 2\delta}{2\delta\sin(\delta -\alpha_{g})\sin(\delta +\alpha_{g})}\Big],\nonumber 
\end{align}
from (\ref{37}). Eq.~(\ref{54}) agrees with the expression for uniform vertical gradient (\ref{45}) if we replace $(G_{x}+G_{y})$ by $-\langle\partial B_{z}/\partial z\rangle$, as justified by (\ref{49}), and use our definition $\zeta=(R/2 B_{0})(\partial B_{z}/\partial z)$. We have shown that this equivalence holds not only in the non-adiabatic limit but throughout the entire range of particle velocities.

To take into account a slight tilt of the Larmor field we could add, in Eq.~(\ref{46}), any small uniform in-plane static field $\boldsymbol{B}_{\rho 0}=(B_{x0}$, $B_{y0})$ without affecting the $E$-odd frequency shift (\ref{54}). The additional terms have the form $B_{\rho 0}\sin\xi$ or $B_{\rho 0}\cos\xi$ and average to zero for random distribution of $\xi$. As a result, up to second order perturbation the $E$-odd frequency shift $\langle\delta\omega\rangle$ from (\ref{54}) is determined solely by the $z$-component $B_{0}$ of the static field.

Along the same lines we can also determine the second-order gradient shift as a function of $\Omega$ and, if desired for better field modeling, extend the polynomial for $\chi(x,y,z)$ in (\ref{47}) to orders $>\!2$ to include non-uniform field gradients. The necessary integrations over $\xi$ and those required for Eqs.~(\ref{19}) can be performed analytically for any order. Only final averaging over $\alpha_{g}$ requires numerical integration in general, but can also be done analytically in the limits $\Omega>>1$ and $\Omega<<1$. In Appendix \ref{sec:A} we extend macroscopic field modeling to fourth order.

\section{Point magnetic dipole on the cylinder axis}\label{sec:V}

Harris and Pendlebury \cite{HAR02} have shown analytically and numerically, using simulations, that the frequency shifts do not necessarily scale with volume-averaged gradient $\langle\partial B_{z}/\partial z\rangle$ for arbitrary magnetic field distributions. For a vertical magnetic dipole on the axis below the floor of a cylindrical cell they obtained significant enhancement in the non-adiabatic limit $\Omega\rightarrow\infty$ (applying to comagnetometer atoms) whereas no enhancement was expected for $\Omega<<1$ (UCNs). Pignol and Roccia \cite{PIG01} then showed that use of the exact dipole field instead of the approximation made in \cite{HAR02} gives essentially the same non-adiabatic enhancement, and extended this work also to general positions on the floor and to horizontal dipole orientation. 

In cylindrical coordinates ($\rho,\phi,z$) the magnetic field of a vertical point dipole placed on the $z$-axis is given by 
\begin{equation}\label{55}
B_{\rho}=\frac{3 p Z}{r^{5}}\rho,\,\,\,B_{\phi}=0,\,\,\, B_{z}=\frac{p}{r^{5}}(3 Z^2-r^{2}),
\end{equation}
where $p$ is the dipole strength, $\boldsymbol{\rho}=(x,y)=(R\cos\alpha_{g},R\Omega\tau)$ and $Z$ denote the horizontal and vertical displacement from the dipole, and $r=\sqrt{Z^{2}+R^{2}\cos^{2}\alpha_{g}+(R\Omega\tau)^{2}}$ is its 3D separation from a point on a path segment as that shown in Fig.~\ref{fig:one}. For the exact dipole field (\ref{55}) the integrations (\ref{19}) cannot be performed analytically. Therefore, we use the approximation, for $Z>0$,
\begin{align}\label{56}
&\mathbf{B} = \nabla\times\mathbf{A};\,\mathbf{A}=-\frac{3p\hat{\boldsymbol{\phi}}}{2\rho^2}\Big[\frac{\rho Z}{\rho^2+Z^2}-\arctan{\frac{\rho}{Z}}\Big];\nonumber \\
& B_{\rho}(\rho,Z)=\frac{3p\rho}{(Z^2+\rho^2)^2};\\
&B_{z}=\frac{3p}{2\rho^3}\Big[\rho Z\frac{3\rho^{2}+Z^{2}}{(\rho^2+Z^2)^{2}}-\arctan{\frac{\rho}{Z}} \Big];\nonumber \\
&\frac{\partial B_{z}}{\partial z}=-\frac{1}{\rho}\frac{\partial}{\partial \rho}[\rho B_{\rho}(\rho,Z)]=-\frac{6p(Z^2-\rho^2)}{(Z^2+\rho^2)^3},\nonumber
\end{align}
with unit vector $\hat{\boldsymbol{\phi}}$ in the azimuthal direction, $Z^2+\rho^2=L^2+(R\Omega\tau)^2$ and $L^{2}=Z^{2}+R^{2}\cos^{2}\alpha_{g}$. Expression (\ref{56}) reduces to (\ref{55}) in the limit $Z>>\rho$, but similar to the approximation $B_{\rho}\simeq 3p\rho Z^{-1}(Z^2+\rho^2)^{-3/2}$ made in \cite{HAR02}, it should be adequate even for a dipole close to or on the cell floor, since the frequency shifts are averages over the cell volume and fairly insensitive to the details of the field model.

We will analyze only the geometric shift linear in $E$ and label terms with subscripts even/odd, depending on whether they are symmetric or antisymmetric under forward/backward transformation ($\Omega\rightarrow -\Omega$, $\alpha_{g}\rightarrow -\alpha_{g}$, $\tau\rightarrow \tau$ \cite{PEN01}). Only the even terms of the frequency shift contribute to the false EDM signal.

Using (\ref{56}) in (\ref{10}) we obtain for $\Sigma_{B}(\tau)=\omega_{0}(B_{x}+iB_{y})/B_{0}=3\omega_{0}p(x+iy)/(B_{0}r^{4})$:   
\begin{align}\label{57}
&\Sigma_{B,\textrm{even}}(\tau)=\frac{M\omega_{0}\cos\alpha_{g}}{\left(w^{2}+\tau^{2}\right)^{2}}, \nonumber \\
&\Sigma_{B,\textrm{odd}}(\tau)=\frac{i\omega_{0}M\Omega\tau}{\left(w^{2}+\tau^{2}\right)^{2}},
\end{align}
where $M=3p/\left(B_{0}R^{3}\Omega^{4}\right)$ and $w=L/(R\Omega)$.

The integration in (\ref{9}), (\ref{10}) gives
\begin{align}\label{58}
&\Sigma_{iB,\textrm{even}}(\tau)=\frac{M\cos\alpha_{g}}{4w^{3}}\textrm{e}^{-i\tau}\Big[\frac{2w\tau}{w^{2}+\tau^{2}}+i(1+w)\,\,\times\nonumber \\
&\textrm{e}^{-w+i\tau}\operatorname{Ei}(w-i\tau)+i(1-w)\textrm{e}^{w+i\tau}\operatorname{E_{1}}(w+i\tau) \Big],\nonumber \\
&\Sigma_{iB,\textrm{odd}}(\tau)=-\frac{iM\Omega}{4w}\textrm{e}^{-i\tau}\Big[\frac{2w}{w^{2}+\tau^{2}}\nonumber \\
&-\textrm{e}^{-w+i\tau}\operatorname{Ei}(w-i\tau)-\textrm{e}^{w+i\tau}\operatorname{E_{1}}(w+i\tau) \Big],
\end{align}
where $\operatorname{Ei}(z)=-\textrm{PV}\!\int_{-z}^{\infty}(\textrm{e}^{-t}/t)dt$ and $\operatorname{E_{1}}(z)=\int_{z}^{\infty}(\textrm{e}^{-t}/t)dt$ are exponential integrals with complex argument (\cite{ABZ01}, 5.1.1, 5.1.2). We use $\operatorname{Ei}$ and $\operatorname{E_{1}}$ in different regions of the complex plane to avoid discontinuities on the time axis due to branch cuts.

The additional integrations required in (\ref{19}) can be performed analytically using the indefinite integrals $\int \textrm{e}^{z}\operatorname{Ei}(-z)dz=-\ln z+\textrm{e}^{z}\operatorname{Ei}(-z)$, $\int \textrm{e}^{z}\operatorname{E_{1}}(z)dz=\ln z+\textrm{e}^{z}\operatorname{E_{1}}(z)$ (\cite{GR01}, 5.231.1).

Proceeding as in sections \ref{sec:II.B}, \ref{sec:II.C} we obtain analytic expressions for $\nu(\delta,-\delta)$ and $s^{2}_{\mu}(\delta,-\delta)$ and the $E$-odd, forward/backward symmetric frequency shift (\ref{36}), (\ref{37})
\begin{align}
&\left\langle\frac{\delta\omega}{\omega_{0}}\right\rangle^{E\rightarrow -E}_{\textrm{even}}=\frac{\left(\nu(\delta,-\delta) \right)_{\textrm{even}}}{\delta}\nonumber \\
&+\frac{\left(s^{2}_{\mu}(\delta,-\delta) \right)_{\textrm{even}}}{n\delta}\sum_{k=1}^{n-1}(n-k)\cos{2 k\alpha_{g}}\sin{2 k\delta}\nonumber \\
&-\frac{\left(s^{2}_{\mu}(\delta,-\delta) \right)_{\textrm{odd}}}{n\delta}\sum_{k=1}^{n-1}(n-k)\sin{2 k\alpha_{g}}\cos{2 k\delta}\label{59} \\
&\stackrel{n>>1}{\longrightarrow} \frac{\left(\nu(\delta,-\delta) \right)_{\textrm{even}}}{\delta}-\frac{\left(s^{2}_{\mu}(\delta,-\delta) \right)_{\textrm{even}}}{2\delta}\left(\cot(\alpha_{g}-\delta)\right)_{\textrm{even}}\nonumber \\
&-\frac{\left(s^{2}_{\mu}(\delta,-\delta) \right)_{\textrm{odd}}}{2\delta}\left(\cot(\alpha_{g}-\delta)\right)_{\textrm{odd}},\label{60}
\end{align}
where
\begin{align}\label{61}
&\left(\cot(\alpha_{g}-\delta)\right)_{\textrm{even}}=\frac{\sin 2\delta}{2\sin(\alpha_{g}-\delta)\sin(\alpha_{g}+\delta)},\nonumber \\
&\left(\cot(\alpha_{g}-\delta)\right)_{\textrm{odd}}=\frac{\sin 2\alpha_{g}}{2\sin(\alpha_{g}-\delta)\sin(\alpha_{g}+\delta)}.
\end{align}

Finally, only averaging over $\alpha_{g}$ and $Z$ has to be performed numerically. For a dipole on the cell floor the integrations include the singular point at $\rho=0$, $Z=0$ but all final integrals remain finite, as they are for the exact dipole field \cite{PIG01}.

\begin{figure}[tb]
  \begin{center} 
\includegraphics[width=42mm]{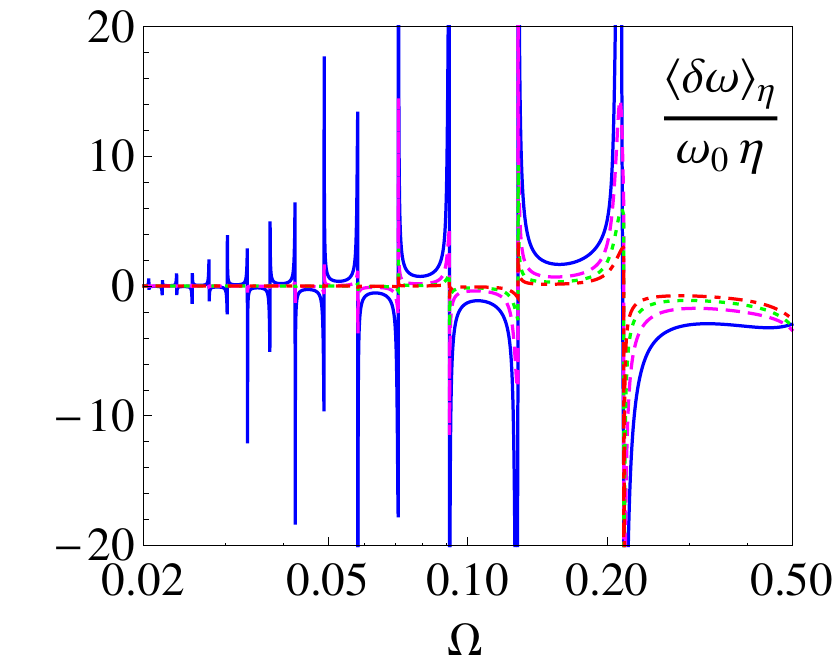}\hspace{1mm}%
 \includegraphics[width=42mm]{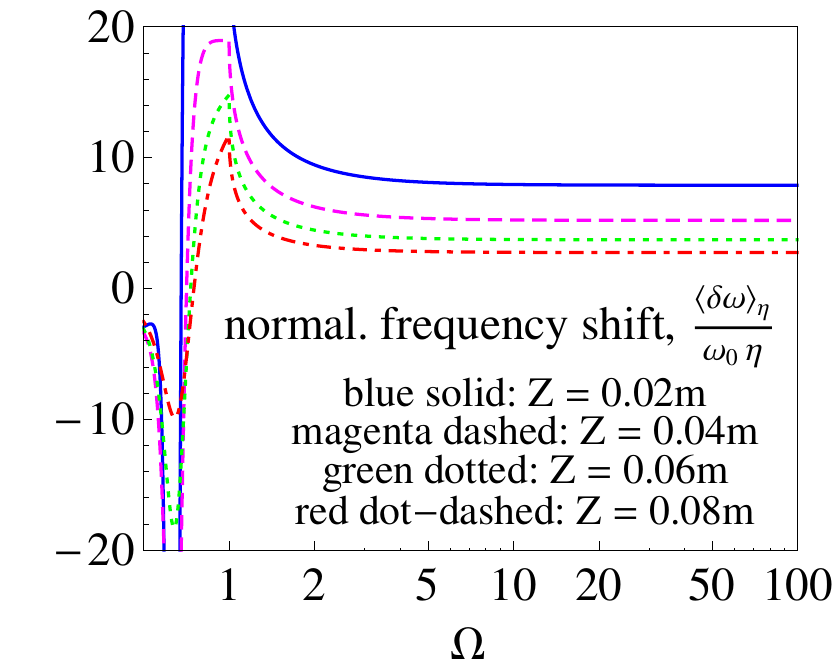}\hspace{40mm}%
  \includegraphics[width=43mm]{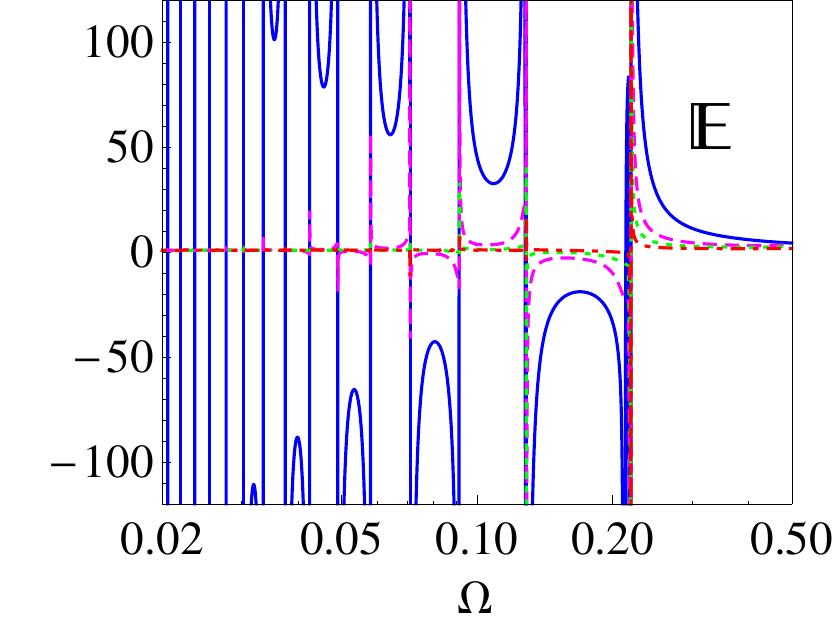}\hspace{0mm}
   \includegraphics[width=42mm]{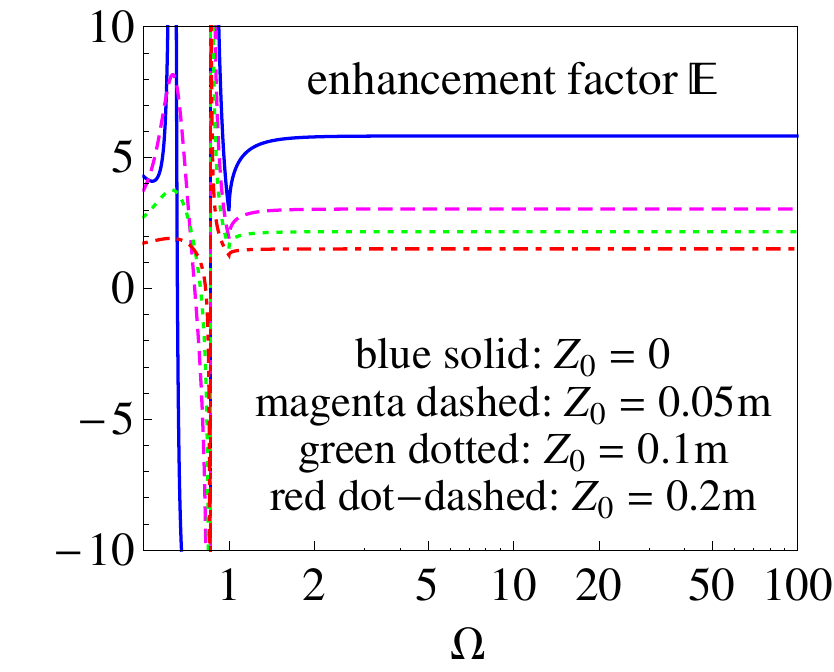}%
 \end{center}
\caption{(Color online)  In two regions of $\Omega$, the upper panels show the normalized frequency shift linear in $E$, $\langle\delta\omega\rangle_{\eta}/(\omega_{0}\eta)$, for particles in horizontal planes at various vertical distances $Z$ above a vertical point dipole of strength $p=R^{3}B_{0}/3$. In the lower panels we plot, in the same $\Omega$ ranges, the enhancement factor $\mathbb{E}$, compared to the shift for uniform vertical gradient $\partial B_{z}/\partial z$. The cell dimensions are those of the ILL experiments, radius $R=0.235$ m, height $H=0.12$ m, and the dipole is located on the cylinder axis a vertical distance $Z_{0}$ below the cell floor.}
  \label{fig:four}
\end{figure}

The upper panels of Fig.~\ref{fig:four} show, in two regions of $\Omega$, the normalized asymptotic frequency shift linear in $E$, $\langle\delta\omega\rangle_{\eta}/(\omega_{0}\eta)=\frac{1}{\eta}\left\langle\frac{\delta\omega}{\omega_{0}}\right\rangle^{E\rightarrow -E}_{\textrm{even}}$, for motion in horizontal planes $Z=0.02$ m, $0.04$ m, $0.06$ m and $0.08$ m above a vertical dipole of strength $p=R^{3}B_{0}/3$. For $\Omega\rightarrow\infty$ the curves approach the non-adiabatic limit which is determined by \cite{PIG01}
\begin{equation}\label{62}
\langle\rho B_{\rho}\rangle_{\rho}=-\frac{3 p}{R^{2}}\left(\frac{R^{2}}{R^{2}+Z^{2}}+\ln\frac{Z^{2}}{R^{2}+Z^{2}} \right),
\end{equation}
the average of $\rho B_{\rho}$ over a horizontal plane, evaluated for dipole model (\ref{56}). For $Z>>R=0.235$ m the calculation, not shown in Fig.~\ref{fig:four}, agrees with the uniform gradient result, as expected since the field approaches that of the uniform vertical gradient. The low-$\Omega$ behavior, shown on the upper left panel, is complex due to resonances at $\Omega \simeq 0.22$, $0.12$, $0.09$, $0.07$, $0.06$, etc.

The lower panels of Fig.~\ref{fig:four} show, in the same $\Omega$ ranges, the enhancement factor \cite{HAR02,PIG01} $\mathbb{E}(\Omega)$, defined as the shift (\ref{59}), (\ref{60}), averaged over $\alpha_{g}$ and $Z$, divided by the shift from (\ref{36}), (\ref{37}) for uniform gradient, with $\zeta=(R/2B_{0})\langle\partial B_{z}/\partial z\rangle$. For our approximation (\ref{56}) the volume averaged gradient is
\begin{align}\label{63}
&\left\langle\frac{\partial B_{z}}{\partial z}\right\rangle=\frac{3p}{H R^{3}}\Big[\frac{RZ_{0}}{R^{2}+Z^{2}_{0}}-\frac{R(Z_{0}+H)}{R^{2}+(Z_{0}+H)^{2}}\nonumber \\
&+\arctan{\frac{Z_{0}}{R}}-\arctan{\frac{Z_{0}+H}{R}}\Big],
\end{align}
with cell height $H$.

In the limit $\Omega\rightarrow\infty$, $\mathbb{E}(\Omega)$ assumes the analytic form \cite{PIG01}
\begin{equation}\label{64}
\mathbb{E}(\Omega\rightarrow\infty)=-1-\frac{4}{R^{2}}\frac{\langle\rho B_{\rho}\rangle_{\rho Z}}{\left\langle\frac{\partial B_{z}}{\partial z}\right\rangle}
\end{equation}
where
\begin{align}\label{65}
&\langle\rho B_{\rho}\rangle_{\rho Z}=\frac{3 p}{R H}\Big[\tan^{-1}\frac{Z_{0}+H}{R}-\tan^{-1}\frac{Z_{0}}{R}\nonumber \\
&-\frac{Z_{0}+H}{R}\ln\frac{(Z_{0}+H)^{2}}{R^{2}+(Z_{0}+H)^{2}}+\frac{Z_{0}}{R}\ln\frac{Z^{2}_{0}}{R^{2}+Z^{2}_{0}} \Big]
\end{align}
is the average of (\ref{62}) over $Z$ for cell height $H$.

The non-adiabatic limit (\ref{64}) of $\mathbb{E}(\Omega)$ for a dipole on the floor ($Z_{0}=0$) is $\mathbb{E}(\infty)=5.8$, quite similar to the value $9.0$ for the exact dipole field \cite{PIG01}. For $Z_{0}=0.05$ m, $0.1$ m, $0.2$ m and $0.3$ m we obtain $\mathbb{E}=3.0$, $2.2$, $1.5$ and $1.3$, to be compared with \cite{PIG01} $4.2$, $2.6$, $1.8$ and $1.4$ and to somewhat larger values for the model of \cite{HAR02}. As seen on the lower right panel of Fig.~\ref{fig:four} there is little dependence of $\mathbb{E}(\Omega)$ on $\Omega$ throughout the high velocity region $\Omega \ge 10$ typical of comagnetometer atoms.

The low-$\Omega$ range of $\mathbb{E}$ shown on the lower left panel ($0.02\le \Omega \le 0.5$) includes the UCN region centered at $\Omega\simeq 0.05$. $\mathbb{E}$ changes sign in consecutive resonance intervals and, especially for $Z_{0}=0$ (blue solid curves with large positive or negative values between resonances), averaging over an actual UCN spectrum would be difficult since the UCN spectra used in the experiments are quite narrow, have a fairly sharp lower and upper cut-off and depend on vertical position due to gravity. For comparison, the adiabatic prediction is a constant enhancement factor $\mathbb{E}(\Omega<<1)=1$ \cite{PEN01,BAR01,HAR02}.

The difference may be due to failure of the adiabatic approximation made in the earlier work. For the point dipole, even slow-moving particles passing close by the dipole in trajectories with $\alpha_{g}\simeq \pi/2$ where most of the shift happens, see a rapidly varying field.

On the basis of Eq.~(\ref{59}) we have also calculated $\mathbb{E}(\Omega)$ for a finite number $n$ of initial wall reflections and found that off resonance the asymptotic values are reached, within $\simeq\!2\%$, after $\simeq\!50-100$ reflections, which is similar to the corresponding number $n\simeq\!20-50$ for the uniform gradient field.

\section{Summary and Conclusions}\label{sec:VI}

Previously, two methods had been used to investigate the geometric phases and frequency shift mimicking a genuine EDM in experiments with confined ultracold neutrons and comagnetometer atoms based on the Ramsey separated oscillatory field magnetic resonance technique: integration of the Bloch equation in \cite{PEN01}, and the Redfield method based on the spin density matrix, which was applied in \cite{LAM01,BAR01} to the EDM system and, in \cite{SWA01}, also to study the general statistical behavior of particles subject to arbitrary fluctuating fields in bounded geometries. Both methods yielded identical results for the frequency shifts in EDM experiments, and this fact was considered ``interesting given the different assumptions made in the two approaches'' \cite{LAM01}. Major assumptions of the Redfield theory are that the time must be short enough so that the evolution of the density matrix is negligible, but long compared to the correlation time.

We have used a third method: direct solution of the Schr\"odinger equation for cylindrical cell geometry to second order in the perturbation. Except for the latter restriction, this method does not rely on any approximations, such as the requirement for the perturbation Hamiltonian $\mathcal{H}_{1}(t)$ of the Redfield theory to have time-average zero, or for the solutions to be stationary. In those cases where comparison is possible (the velocity dependence of shifts in a vertical gradient field \cite{PEN01,LAM01,BAR01} and the non-adiabatic limit for general magnetic gradient and for dipole fields \cite{HAR02,PIG01}) we have obtained results identical to those of the earlier studies.

As new elementa, our analysis has allowed us to study also the non-stationary, transient spin behavior, i.e., the gradual development of the shifts for an arbitrary number $n$ of wall reflections subsequent to the start of the period of free spin precession following the first $\pi/2$ pulse in the Ramsey scheme. Our general solution, described in Sec.~\ref{sec:II}, also provides full information on the vertical spin oscillations associated with the phase shifts, for arbitrary particle velocity and arbitrary number $n$ of successive reflections. For the general uniform and non-uniform gradient fields analyzed in Sec.~\ref{sec:IV} and in the Appendix, and for the field of a vertical magnetic dipole on the cylinder axis (in Sec.~\ref{sec:V}), where so far analytic expressions had been known \cite{PIG01} only for the non-adiabatic limit of frequency shift (i.e., for particle velocity $v\rightarrow\infty$), we have obtained analytical results valid for any velocity.

%
\acknowledgments
A. S. acknowledges travel support by the UCN group of the Los Alamos National Laboratory and by the Excellence Cluster Universe of the Technical University M\"unchen. 

\appendix
 
\section{Fourth-order expansion of the magnetic field}\label{sec:A}
 
When we extend the magnetic potential (\ref{47}) to include $3^{\textrm{rd}}$ and $4^{\textrm{th}}$ order terms we note that there are 9 third-order terms $Q_{ijk}x_{i}x_{j}x_{k}$ where each of the indices $i$, $j$, $k$ can represent $x$, $y$ or $z$. With three constraints imposed by $\nabla^{2}\chi(x,y,z)=0$ on the $3^{\textrm{rd}}$ order terms we have $6$ independent coefficients $Q_{ijk}$ characterizing expansion terms such as $Q_{xxz}x^2z$. Similarly, there are $15$ fourth-order coefficients $Q_{ijkl}$ and $6$ additional constraints, thus $9$ independent fourth-order contributions $Q_{ijkl}x_{i}x_{j}x_{k}x_{l}$ to the magnetic potential, e.g. $Q_{yyzz}y^2z^2$.

In terms of these coefficients, the first-order and third-order vertical field gradients $\partial B_{z}/\partial z$ and $\partial^{3} B_{z}/\partial z^{3}$, averaged over cell volume, are
\begin{align}\label{A1}
&\left\langle\frac{\partial B_{z}}{\partial z}\right\rangle=-(G_{x}+G_{y})+\frac{3R^{2}-H^2}{6}(Q_{xxzz}+Q_{yyzz}),\nonumber \\
&\left\langle\frac{\partial^{3} B_{z}}{\partial z^{3}}\right \rangle=-4(Q_{xxzz}+Q_{yyzz}),
\end{align}
where we have used the volume averages $\langle x^{2}\rangle=\langle y^{2}\rangle=R^{2}/4$, $\langle z^{2}\rangle=H^{2}/12$ and $\langle x\rangle=\langle y\rangle=\langle z\rangle=0$. Only the two quantities in (\ref{A1}) will be relevant below.

Proceeding as in Sec.~\ref{sec:V}, Eqs.~(\ref{48})-(\ref{54}), and neglecting (as in \cite{PIG01}) the slight dependence of particle spectra on vertical position within the cell due to gravity, we obtain for the geometric frequency shift for $E$-field reversal
\begin{align}\label{A2}
&\left\langle\frac{\delta\omega}{\omega_{0}} \right\rangle^{E\rightarrow -E}=\nonumber \\
&-\frac{\eta R\Omega^{2}}{B_{0}}(G_{x}+G_{y})\Big[1+\frac{\sin^{2}\alpha_{g}\sin 2\delta}{2\delta\sin(\delta -\alpha_{g})\sin(\delta +\alpha_{g})}\Big]\nonumber\\
&+\frac{\eta R\Omega^{2}(Q_{xxzz}+Q_{yyzz})}{6 B_{0}}  \Big\{3 R^{2}(1-4\Omega^{2})-H^{2}+\nonumber \\
&\frac{1}{2\delta\sin(\delta -\alpha_{g})\sin(\delta +\alpha_{g})}\Big[6R^{2}\Omega^{2}\delta\left(1-\cos 2\delta \cos 2\alpha_{g}\right)\nonumber \\
&+[9R^{2}(1-2\Omega^{2})-H^{2}]\sin^{2}\alpha_{g}\sin 2\delta \Big]\Big\}\nonumber \\
&=\frac{\eta R\Omega^{2}}{B_{0}}\left\langle\frac{\partial B_{z}}{\partial z}\right\rangle\Big[1+\frac{\sin^{2}\alpha_{g}\sin 2\delta}{2\delta\sin(\delta -\alpha_{g})\sin(\delta +\alpha_{g})}\Big]\nonumber \\
&-\frac{\eta R^{3}\Omega^{3}\sin\alpha_{g}}{2 B_{0}}\left\langle\frac{\partial^{3} B_{z}}{\partial z^{3}}\right\rangle \Big\{\frac{(1-3\Omega^{2})\sin 2\delta}{4\sin(\delta -\alpha_{g})\sin(\delta +\alpha_{g})}\nonumber \\
&-\frac{1}{\delta}\Big[1-\frac{\sin(\delta-\alpha_{g})}{4\sin(\delta+\alpha_{g})}-\frac{\sin(\delta+\alpha_{g})}{4\sin(\delta-\alpha_{g})} \Big] \Big\}. 
\end{align}

The second and sixth lines in (\ref{A2}) are identical to Eq.~(\ref{54}) for quadratic $\chi(x,y,z)$ where $G_{x}+G_{y}=-\langle\partial B_{z}/\partial z\rangle$ is directly determined by the volume-averaged first-order vertical gradient of $B_{z}$. The remainder represents the correction due to $\langle\partial^{3} B_{z}/\partial z^{3}\rangle$ from (\ref{A1}). It could be the dominant contribution to the frequency shift in measurement cells with small first-order gradient but non-negligible higher-order inhomogeneities.

In Eq.~(\ref{A2}), as in (\ref{43})-(\ref{45}), the terms $\propto\!\delta^{-1}$ are dominant in the non-adiabatic limit ($\Omega>>1$, $\delta<<1$) and become negligible in the adiabatic limit ($\Omega<<1$, $\delta>>1$). We average (\ref{A2}) over $\alpha_{g}$ and obtain the following expansion for $\Omega>>1$:
\begin{align}\label{A3}
&\left\langle\left\langle\frac{\delta\omega}{\omega_{0}} \right\rangle\right\rangle^{E\rightarrow -E}_{\alpha_{g}}=-\frac{\eta R}{2 B_{0}}\times\\
&\Big[\left\langle\frac{\partial B_{z}}{\partial z}\right\rangle +\frac{R^{2}}{24}\left\langle\frac{\partial^{3} B_{z}}{\partial z^{3}}\right\rangle \Big]\left(1+\frac{2}{3\Omega^{2}}\right)+O[\Omega^{-4}].\nonumber 
\end{align}

In the non-adiabatic limit $\Omega\rightarrow\infty$, Eq.~(\ref{A3}) agrees with the general result from [\cite{PIG01}, Eq.~(5)],
\begin{equation}\label{A4}
\langle\delta\omega\rangle^{E\rightarrow -E}=2\frac{\gamma^{2}E}{c^{2}}\langle\rho B_{\rho} \rangle=\frac{2\eta\omega_{0}}{R B_{0}}\langle\rho B_{\rho} \rangle,
\end{equation}
which is determined by the volume average
\begin{align}\label{A5}
&\langle\rho B_{\rho}\rangle =\frac{R^{2}}{4}\Big[G_{x}+G_{y}-\frac{2 R^{2}-H^{2}}{6}(Q_{xxzz}+Q_{yyzz})\Big]\nonumber \\
&=-\frac{R^{2}}{4}\Big[\left\langle\frac{\partial B_{z}}{\partial z}\right\rangle+\frac{R^{2}}{24}\left\langle\frac{\partial^{3} B_{z}}{\partial z^{3}}\right\rangle\Big],
\end{align}
with gyromagnetic ratio $\gamma=2\mu/\hbar$. Our result agrees with the sign in \cite{PIG01} taking into account a different sign convention. For positive gradient $\left\langle\partial B_{z}/\partial z\right\rangle$ (and neglecting $\left\langle\partial^{3} B_{z}/\partial z^{3}\right\rangle$) we have for positive magnetic moment $\mu$, as for $^{199}$Hg: $\omega_{0}<0$ and $\langle\delta\omega\rangle/\omega_{0}<0$, thus a decrease of the magnitude of precession frequency. The latter is true also for negative $\mu$, as for $^{3}$He or $^{129}$Xe, for which $\omega_{0}>0$ and $\langle\delta\omega\rangle/\omega_{0}<0$.

In the adiabatic range $\Omega<<1$ we obtain, after averaging over the resonances,
\begin{align}\label{A6}
&\left\langle\left\langle\frac{\delta\omega}{\omega_{0}} \right\rangle\right\rangle^{E\rightarrow -E}_{\alpha_{g}}\\
&=\frac{\eta R\Omega^{2}}{B_{0}}\Big[\left\langle\frac{\partial B_{z}}{\partial z}\right\rangle +\frac{3 R^{2}\Omega^{2}}{4}\left\langle\frac{\partial^{3} B_{z}}{\partial z^{3}}\right\rangle \Big],\nonumber
\end{align}
which is almost exclusively determined by the first-order gradient $\langle\partial B_{z}/\partial z\rangle$. In this range, i.e. for UCNs, $\omega_{0}>0$ and the relative frequency change is opposite to that for $\Omega>>1$, i.e. the magnitude of precession frequency increases for $\left\langle\partial B_{z}/\partial z\right\rangle>0$, and decreases for $\left\langle\partial B_{z}/\partial z\right\rangle<0$.

\end{document}